\documentclass[USletter,10pt]{article}
\usepackage[margin=1in,dvips]{geometry}
\usepackage{graphicx,cancel}
\usepackage{graphics}
\usepackage{color, soul}
\usepackage{amsmath, amsthm, slashed}
\usepackage{amssymb}
\usepackage{rotating}
\usepackage{mathrsfs}
\usepackage{epsfig}
\usepackage{verbatim}
\usepackage[affil-it]{authblk}
\usepackage{rotating}
\usepackage{graphicx}
\usepackage{accents}

\newtheorem{theorem}{Theorem}[section]
 
\newtheorem*{conjecture*}{Conjecture}
\newtheorem{corollary}{Corollary}[section]
\newtheorem*{theorem*}{Theorem}
\newtheorem*{corollary*}{Corollary}

\DeclareMathAlphabet\mathbfcal{OMS}{cmsy}{b}{n}

\title{A note on the wave equation on black hole spacetimes \\ with small non-decaying first order terms}

\author{Gustav Holzegel and Christopher Kauffman}

\affil{\small Imperial College London, Department of Mathematics, 
South~Kensington~Campus,~London~SW7~2AZ,~United~Kingdom\vskip.2pc \ }

\let\oldmarginpar\marginpar
\renewcommand\marginpar[1]{\-\oldmarginpar[\raggedleft\footnotesize #1]%
{\raggedright\footnotesize #1}}

\begin{document}
\maketitle

\abstract{We present an elementary physical space argument to establish local integrated decay estimates for the perturbed wave equation $\Box_g \phi = \epsilon \beta^a \partial_a \phi$ on the exterior of the Schwarzschild geometry $(\mathcal{M},g)$. Here $\beta$ is a regular vectorfield on $\mathcal{M}$ decaying suitably in space but not necessarily in time. The proof is formulated to cover also perturbations of the Regge--Wheeler equation.}

\section{Introduction}
In the past 15 years a rather satisfactory and complete picture of the global behaviour of solutions to the linear wave equation, $\Box_g \phi=0$, on black hole backgrounds has emerged. The results range from decay results in the exterior Schwarzschild and Kerr geometries \cite{partiii, DafRodsmalla, Toha2,AndBlue, DafRod2, MR1972492, Sterbenz, Aretakis} to the understanding of precise asymptotics \cite{Angelopoulos:2016wcv, Angelopoulos:2016moe,Hintz:2020roc} to the analysis of the wave equation on non-stationary spacetimes eventually converging to Schwarzschild or Kerr \cite{tohaneanu, Lindbladglobal2016,Lindblad:2020bkn}. 

Given the wealth of results one may ask the following general question: Which statements continue to hold when a small first order \emph{non-decaying}  (in time) perturbation is added to the equation? More specifically, given the equation
\begin{align} \label{perturbed}
\Box_g \phi = \epsilon \beta^a \partial_a \phi \, ,
\end{align}
where $(\mathcal{M},g)$ is a black hole spacetime and $\beta$ is a regular vectorfield on $\mathcal{M}$ (obeying suitable spatial asymptotics), what can we say about the global behaviour of $\phi$ on $(\mathcal{M},g)$? 

In this short note, we shall restrict ourselves to $(\mathcal{M},g)$ being the Schwarzschild spacetime and provide an elementary, \emph{purely physical space} argument to establish a non-degenerate integrated decay estimate for solutions to (\ref{perturbed}), see Theorem \ref{theo:mtheo}. The method generalises  to the case of the Kerr spacetime, suitably applied to the frequency decomposed parts of the solution. Nevertheless, in order to not distract from the elementary nature of the argument, we will provide the details as well as further geometric considerations in a separate paper.

To appreciate the difficulty in establishing integrated local energy decay (ILED) estimates for (\ref{perturbed}) in the black hole setting, we first observe that if $g$ is the Minkowski metric in (\ref{perturbed}), then we have, under appropriate pointwise assumptions on $\beta$, the well-know estimates
\begin{align}
\int_{\Sigma_t} d^3 x |D\phi|^2 &\lesssim  \int_{\Sigma_0} d^3 x |D\phi|^2 + \epsilon \int_0^t d\tau \int_{\Sigma_{\tau}} d^3 x \frac{1}{1+|x|^2} |D\phi|^2  \ \ \ \textrm{energy estimate,}
 \\
 \int_0^t d\tau \int_{\Sigma_{\tau}} d^3 x \frac{1}{1+|x|^2} |D\phi|^2 &\lesssim \int_{\Sigma_0} d^3 x |D\phi|^2 + \epsilon \int_0^t d\tau \int_{\Sigma_{\tau}} d^3 x \frac{1}{1+|x|^2} |D\phi|^2 \ \ \ \ \textrm{ \ \ \ \ ILED,}   \label{ILED}
\end{align}
which can be immediately coupled to produce an ILED for $\phi$ in terms of the initial energy. In the above, $\Sigma_t$ denotes a hypersurface of constant $t$ and $|D\phi|^2 = (\partial_t \phi)^2 + |\nabla \phi|^2$ the standard energy density.

In sharp contrast, already in the Schwarzschild case, even for the free wave equation we can --  at the level of first derivatives -- only bound a \emph{degenerate} (near $r=3M$) spacetime integral in terms of the initial energy \cite{SbierskiGauss}.\footnote{See the energy $\mathbb{I}_{deg}\left[\phi\right](\tau_1,\tau_2)$ defined in Section \ref{sec:energies} for the energy that can be controlled by the initial energy.} This analytic fact originates in the well-known geometric phenomenon of trapped null geodesics, which in the Schwarzschild geometry concentrate on the timelike hypersurface $r=3M$. As a consequence, because in the perturbed problem (\ref{perturbed}) the $\epsilon$-term in (\ref{ILED}) will still be \emph{non-degenerate} (unless $\beta$ is assumed to degenerate at $r=3M$) we cannot absorb the error on the left and the estimates do not close. Note also that commuting by the symmetries of Schwarzschild and trying to close the estimates at higher orders does not improve the situation. In other words, one cannot treat the error perturbatively using the standard techniques.

The resolution, in a nutshell, is to commute (\ref{perturbed}) by a suitable vectorfield $W$ (defined in (\ref{Wdef}) below) such that one can control, up to lower order errors, \emph{non-degenerately} the spacetime integral involving all first derivatives of $W\phi$. The lower order errors can then in turn be controlled by the \emph{degenerate} spacetime integral for first derivatives plus a small amount of the non-degenerate spacetime integral involving all first derivatives of $W\psi$. Hence at this level, the estimates close.

On a more abstract level, the fact that one can prove decay estimates for the perturbed equation (\ref{perturbed}) is related to the normally hyperbolic (unstable) structure of trapping in Schwarzschild and Kerr spacetimes together with the observation that first order perturbations do not change the geodesic flow associated with the wave operator. See \cite{Wunsch, Dyatlovnh} for some general results on semi-classical resolvent estimates for the wave operator in the presence of normally hyperbolic trapping including first order perturbations \cite{Hintznh}.

We shall postpone further comments about the relation of (\ref{perturbed}) with the stability of black holes to Section \ref{sec:finalcomments} after having introduced some notation.

\section{Preliminaries}
Fix $M>0$. We recall the Schwarzschild metric written in regular coordinates $(t,r,\theta,\phi)$ as
\begin{align}
g= -\left(1-\frac{2M}{r}\right)dt^2 + \frac{4M}{r} dt dr + \left(1+\frac{2M}{r}\right) dr^2 + r^2 \left(d\theta^2 +\sin^2 \theta d\varphi^2\right) \,  \, ,
\end{align}
defined on the (exterior) manifold $\mathcal{M}=\left(-\infty,\infty\right)\times [2M,\infty) \times S^2$. Its boundary $\mathcal{H}^+ = (-\infty,\infty) \times \{r=2M\} \times S^2$ is a null hypersurface, the event horizon. Slices of constant $t=\tau$ in $\mathcal{M}$ are spacelike and denoted $\Sigma_\tau$. We also define for $\tau_1<\tau_2$ the spacetime region $\mathcal{M}(\tau_1,\tau_2)=(\tau_1,\tau_2) \times [2M,\infty) \times S^2$. 

Note the vectorfield $T= \partial_t$ is Killing for $g$ and null on $\mathcal{H}^+$. We introduce the vectorfield\footnote{The notation is reminiscent of the fact that in the irregular tortoise coordinates $(t,r^\star,\theta,\phi)$, we have $R^\star=\partial_{r^\star}$.} $R^\star=\frac{2M}{r}\partial_{t} + (1-\frac{2M}{r})\partial_r$ and the vectorfield
\begin{align} \label{Wdef}
W := \frac{r}{\sqrt{1-\frac{2M}{r}}}\left( R^\star + f(r) T\right) \ \ \ \textrm{with} \ \ \ f(r)=  \left(1-\frac{3M}{r}\right)\sqrt{1+\frac{6M}{r}} \, ,
\end{align}
whose relevance will become clear below. Note already that the vectorfield $\left(1-\frac{2M}{r}\right)^{-\frac{1}{2}}W$ extends regularly to the horizon and that $f=1+\mathcal{O}(r^{-2})$ near infinity.

It will be convenient to reformulate (\ref{perturbed}). We define for $s=0,1,2$ the operator
\begin{align} \label{psintro}
\mathcal{P}_s \psi := \left(1+\frac{2M}{r}\right)\partial_t^2 \psi -\frac{4M}{r} \partial_t \partial_r \psi + \frac{2M}{r^2}\partial_t \psi -\partial_r \left(\left(1-\frac{2M}{r}\right)\partial_r \psi \right) -\frac{\Delta_{S^2}\psi}{r^2}+ (1-s^2)\frac{2M}{r^3} \psi \, ,
\end{align}
where $\Delta_{S^2}$ denotes the Laplacian on the unit sphere. Note that if $\phi$ satisfies (\ref{perturbed}), then $\psi=\phi \cdot r$ will satisfy $\mathcal{P}_0 \psi =  \epsilon  \left( \beta^a \partial_a \psi - \frac{\beta(r)}{r} \psi\right)$. We have introduced the parameter $s$ since the operator $\mathcal{P}_s$ is closely related to the Regge-Wheeler operator\footnote{\label{ftr} This is perhaps more manifest by expressing the operator in the irregular tortoise coordinates $(t,r^\star,\theta,\phi)$ where one has $
(1-\frac{2M}{r})\mathcal{P}_s \psi=\partial_t^2 \psi - (\partial_{r^\star})^2 \psi -\frac{1-\frac{2M}{r}}{r^2} \Delta_{S^2}\psi + (1-s^2)\frac{2M}{r^3} \left(1-\frac{2M}{r}\right)\psi$. In the actual (angular separated) Regge-Wheeler equation it is $\ell(\ell+1)$ with $\ell \geq |s|$ which appears in place of $\Delta_{S^2}$. We achieve the same effect by restricting the data in (\ref{reformulate}) appropriately for $s=1,2$.} 
of spin $s$ and our results will equally apply to perturbed Regge-Wheeler equations, which are important in applications.

Let $\beta$ be a smooth vectorfield on $\mathcal{M}$. We denote by $\slashed{\beta}$ its projection to the tangent space of the spheres of symmetry (constant $t$ and $r$). We assume $\beta$ to satisfy on all of $\mathcal{M}$ the uniform bound
\begin{align} \label{betabounds}
r^2 |\beta^t| + r^2 |\beta^r| + r^\frac{3}{2}\sqrt{\slashed{g}_{AB}\slashed{\beta}^A \slashed{\beta}^B} + r^\frac{3}{2} |W(\beta^t)| +  r^\frac{3}{2} |W(\beta^r)| + r\sqrt{\slashed{g}_{AB}W(\slashed{\beta}^A) W(\slashed{\beta}^B)} \leq C \, ,
\end{align}
where $\slashed{g}$ denotes the (round) metric induced on the spheres of symmetry.

We can finally set up the Cauchy problem we would like to study. We fix the spacelike hypersurface $\Sigma_0$ and consider 
\begin{align} \label{reformulate}
\mathcal{P}_s \psi = \epsilon {\beta}^a \partial_a \psi \ \ \ \ \ ,  \ \ \ \ \psi\Big|_{\Sigma_0}=\psi_0 \ \ \ , \ \ \ n_{\Sigma_0}\psi\Big|_{\Sigma_0}=\psi_1 \, ,
\end{align}
where for simplicity $\psi_0$ and $\psi_1$ are assumed to be smooth functions of compact support on $\Sigma_0$ and
\begin{itemize}
\item in case $s=1$ we assume $\psi_0, \psi_1$ to have vanishing spherical mean, 
\item in case $s=2$ we assume $\psi_0, \psi_1$ to have vanishing projection to the $\ell=0,1$ spherical harmonics. 
\end{itemize}
The additional assumptions for $s=1,2$ are justified by the physical origin of these equations (footnote \ref{ftr}). They imply that the energy associated with the vectorfield $T$ is always coercive.
Note we have deliberately dropped the zeroth order term on the right hand side of (\ref{reformulate}) that arose in the derivation from (\ref{perturbed}) as it provides no difficulty in the analysis and can be easily carried through if desired.

\section{Energies} \label{sec:energies}
We define the following energies for $\psi$, where we write short hand $\int_{\Sigma_\tau}$ for $\int_{\Sigma_\tau} dr \sin \theta d\theta d\phi$.
\begin{align}
\mathbb{E}[\psi](\tau) &:= \int_{\Sigma_{\tau}} (\partial_t \psi)^2 + \left(1-\frac{2M}{r}\right)  (\partial_r \psi)^2 + |\slashed{\nabla} \psi|^2 + \frac{\psi^2}{r^2}  \, ,
\nonumber \\
\mathbb{I}_{deg} [\psi](\tau_1,\tau_2) &:= \int_{\tau_1}^{\tau_2} d\tau \int_{\Sigma_{\tau}} \left(1-\frac{3M}{r}\right)^2 \left(\frac{1}{r^2} (\partial_t \psi)^2+\frac{1-\frac{2M}{r}}{r^2} (\partial_r \psi)^2 + \frac{1}{r} |\slashed{\nabla} \psi|^2\right) + \frac{1}{r^2}|R^\star \psi|^2 + \frac{\psi^2}{r^3} \, , 
\nonumber \\ 
\mathbb{I}_{ndeg} [\psi](\tau_1,\tau_2) &:= \int_{\tau_1}^{\tau_2} d\tau \int_{\Sigma_{\tau}}  \frac{1}{r^2} (\partial_t \psi)^2+\frac{1-\frac{2M}{r}}{r^2} (\partial_r \psi)^2 + \frac{1}{r} |\slashed{\nabla} \psi|^2 + \frac{\psi^2}{r^3}\, . \nonumber
\end{align}
Note that for the above energies, the transversal derivative $\partial_r$ degenerates near the horizon. We also define the non-degenerate energies
\begin{align}
\overline{\mathbb{E}}[\psi](\tau) &:= \int_{\Sigma_{\tau}} (\partial_t \psi)^2 +  (\partial_r \psi)^2 + |\slashed{\nabla} \psi|^2 + \frac{\psi^2}{r^2}  \, , 
\nonumber \\
\overline{\mathbb{I}}_{deg} [\psi](\tau_1,\tau_2) &:= \int_{\tau_1}^{\tau_2} d\tau \int_{\Sigma_{\tau}} \left(1-\frac{3M}{r}\right)^2 \left(\frac{1}{r^2} (\partial_t \psi)^2+\frac{1}{r^2} (\partial_r \psi)^2 + \frac{1}{r} |\slashed{\nabla} \psi|^2\right) + \frac{1}{r^2}|R^\star \psi|^2 + \frac{\psi^2}{r^3}  \, , 
\nonumber \\ 
\overline{\mathbb{I}}_{ndeg} [\psi](\tau_1,\tau_2) &:= \int_{\tau_1}^{\tau_2} d\tau \int_{\Sigma_{\tau}}  \frac{1}{r^2} (\partial_t \psi)^2+\frac{1}{r^2} (\partial_r \psi)^2 + \frac{1}{r} |\slashed{\nabla} \psi|^2 + \frac{\psi^2}{r^3} \, . \nonumber
\end{align}
Throughout this paper we use the notation $X \lesssim Y $ to denote that $X\leq \tilde{C} Y$ holds with a constant $\tilde{C}$ depending only on $M$ and the constant $C$ in (\ref{betabounds}).

\section{Main Theorem}
\begin{theorem} \label{theo:mtheo}
Consider the Cauchy problem (\ref{reformulate}) with the vectorfield $\beta$ satisfying (\ref{betabounds}) on $\mathcal{M}$. Then there exists $\epsilon_0>0$ small (depending only on $M$ and $C$ in (\ref{betabounds})) such that for any $0\leq \epsilon\leq \epsilon_0$ all smooth solutions to (\ref{reformulate}) satisfy for all $\tau>0$ the estimate
\begin{align} \label{mainestimate}
\overline{\mathbb{E}}[\psi](\tau) + \mathbb{E}[W\psi](\tau) + \overline{\mathbb{I}}_{ndeg} [\psi](0,\tau)  + \mathbb{I}_{ndeg} [W\psi](0,\tau) \lesssim \overline{\mathbb{E}}[\psi](0) + \mathbb{E}[W\psi](0) \, .
\end{align}
\end{theorem}

We abstain from proving or even stating higher order estimates as these follow by now well-established techniques \cite{Mihalisnotes}. In particular, we easily obtain the following loosely stated corollary \cite{DafRodnew}:

\begin{corollary}
Under the assumptions of the previous theorem and assuming in addition the boundedness of higher order weighted energies initially, solutions to (\ref{reformulate}) decay inverse polynomially in time on $\mathcal{M}$.
\end{corollary}

We finally remark that for (\ref{reformulate}), unlike for the free wave equation, uniform boundedness of the energy without loss of derivatives or degeneration at $3M$ (i.e.~the estimate $\overline{\mathbb{E}}[\psi](\tau) \lesssim \overline{E}[\psi](0)$) cannot hold in general. This follows directly from Theorem 3.D.2 of \cite{sbierskithesis}.

\section{Proof of the main theorem}
We first note the following non-degenerate (at the horizon) energy boundedness and degenerate (at $r=3M$) integrated local energy decay estimate for $\psi$ itself, which are by now very standard \cite{Mihalisnotes, holzstabofschw}: 
\begin{align} 
\overline{\mathbb{E}}[\psi](\tau) &\lesssim \overline{\mathbb{E}}[\psi](0) + \epsilon \cdot \overline{\mathbb{I}}_{ndeg} [\psi](0,\tau)   \label{basicenergy} \, ,
 \\
\overline{\mathbb{I}}_{deg} [\psi](0,\tau) &\lesssim \overline{\mathbb{E}}[\psi](0) + \epsilon \cdot \overline{\mathbb{I}}_{ndeg} [\psi](0,\tau) \label{basicILED} \, .
\end{align}
The estimates (\ref{basicenergy}) and (\ref{basicILED}) arise from applying to (\ref{reformulate}) a multiplier of the form $f_1 \partial_t \psi + f_2 \partial_r \psi + \frac{1}{r} f_3 \psi$ with suitably chosen bounded radial functions $f_1,f_2,f_3$ and integrating over $\mathcal{M}(0,\tau)$ with the measure $dt dr \sin \theta d\theta d\phi$. Hence  the $\epsilon$-term appearing on the right hand side of (\ref{basicenergy}), (\ref{basicILED})
is easily derived using (\ref{betabounds}).  

The second and main step is to commute (\ref{reformulate}) by the vectorfield $W$ defined in (\ref{Wdef}). A lengthy but entirely straightforward computation yields
\begin{align} \label{regcommuted}
\mathcal{P}_s (W\psi)  = -\frac{2}{r}\frac{1}{\sqrt{1+\frac{6M}{r}}} \frac{1}{1-\frac{2M}{r}}  T(W\psi) +  \mathcal{F}_1 + \mathcal{F}_2
\end{align}
with 
\begin{align}
\mathcal{F}_1 = \epsilon \frac{1}{r^2} W \left[ r^2  \left( {\beta}^a \partial_a \psi \right)\right]
\ \ \ \ , \ \ \ \ 
\mathcal{F}_2 =  \left(1-\frac{2M}{r}\right)^{-1/2} \left( h_{2} \partial_r\psi + h_{2} \partial_t \psi + h_{3} \psi \right)
\end{align}
and where the $h_b$ denote (explicit) bounded functions of $r$, $h_b: \left[2M,\infty\right) \rightarrow \mathbb{R}$ which satisfy the asymptotic behaviour $\sup_{[2M,\infty)} | r^b h_b| \leq \hat{C}$ with $\hat{C}$ depending only on $M$.\footnote{Note the same symbol $h_b$ may represent different explicit functions in different places.} Note that while the right hand side of (\ref{regcommuted}) appears singular near the horizon, there are cancellations between the first term and $\mathcal{F}_2$ which make it regular. However, we will not need to exploit these cancellations to close the estimates. 

Multiplying now (\ref{regcommuted}) by $TW\psi$  and integrating over the region $\mathcal{M}(0,\tau)$ we easily deduce (after applying Cauchy Schwarz on the right using again (\ref{betabounds}) for the terms involving $\beta$)
\begin{align} \label{energyWpsi}
\mathbb{E}[W\psi](\tau) + \int_{0}^{\tau} d\tilde{\tau} \int_{\Sigma_{\tilde{\tau}}} \frac{1}{r} \frac{1}{1-\frac{2M}{r}} (\partial_t W\psi)^2  \lesssim  \mathbb{E}[W\psi](0) + \overline{\mathbb{I}}_{ndeg} [\psi](0,\tau) + \epsilon \cdot \mathbb{I}_{ndeg} [W\psi](0,\tau)  \, .
\end{align}
Note the favourable non-degenerate spacetime term that is being produced. This term has optimal weights near the horizon (and hence a version of the redshift is implicit in the commutation with $W$) and strong $r$-weights near infinity capturing that $W$ is asymptotically tangent to the outgoing null cones, $W\sim \frac{r}{\sqrt{1-\frac{2M}{r}}}(T+R^\star) + \mathcal{O}(r^{-1})\partial_t$.

By a Lagrangian estimate (multiplying (\ref{regcommuted}) by $\frac{1}{r} W\psi$ and integrating over $\mathcal{M}(0,\tau)$) we have
\begin{align} \label{Lagrangian}
\mathbb{I}_{ndeg} [W\psi](0,\tau) \lesssim   \int_{0}^{\tau} d\tilde{\tau} \int_{\Sigma_{\tilde{\tau}}} \frac{1}{r} \frac{1}{1-\frac{2M}{r}} (\partial_t W\psi)^2 &+  \overline{\mathbb{I}}_{ndeg} [\psi](0,\tau) \nonumber \\
&+ \mathbb{E}[W\psi](\tau) + \mathbb{E}[W\psi](0) +  \overline{\mathbb{E}}[\psi](\tau) +  \overline{\mathbb{E}}[\psi](0) \, .
\end{align}
To obtain (\ref{Lagrangian}) note in particular the bounds
\begin{align}
  \int_{\Sigma_\tau} \frac{|W\psi|^2}{1-\frac{2M}{r}} \frac{1}{r^2}   \lesssim \overline{\mathbb{E}}[\psi](\tau)  \ \ \ \textrm{and} \ \ \  \int_{0}^{\tau} d\tilde{\tau} \int_{\Sigma_{\tilde{\tau}}} \frac{|W\psi|^2}{1-\frac{2M}{r}} \frac{1}{r^3}  \lesssim \overline{\mathbb{I}}_{ndeg}[\psi](0,\tau)   \, . \nonumber 
\end{align}
Combining (\ref{energyWpsi}) and (\ref{Lagrangian}) and (\ref{basicenergy}) we deduce
\begin{align} \label{make2}
\mathbb{E}[W\psi](\tau)  + \mathbb{I}_{ndeg} [W\psi](0,\tau) \lesssim \mathbb{E}[W\psi](0) + \overline{\mathbb{E}}[\psi](0)  + \overline{\mathbb{I}}_{ndeg} [\psi](0,\tau) \, .
\end{align}
The estimates will now close  from the following elementary estimate, valid for any $\delta>0$
\begin{align} \label{Hardy}
 \overline{\mathbb{I}}_{ndeg} [\psi](0,\tau) \lesssim \frac{1}{\delta} \overline{\mathbb{I}}_{deg} [\psi](0,\tau) + \delta \cdot \mathbb{I}_{ndeg} [W\psi](0,\tau) + \mathbb{E}[\psi](\tau) + \mathbb{E}[\psi](0) \, .
\end{align}
Indeed, inserting (\ref{make2}) and (\ref{basicILED}) into (\ref{Hardy}) and choosing $\delta$ sufficiently small depending only on $M$ and $C$ in (\ref{betabounds}) we obtain (\ref{mainestimate}). We finally indicate how to prove (\ref{Hardy}). We let $\chi : [2M,\infty) \rightarrow \mathbb{R}$ be a radial cut-off function equal to $1$ in $\left[\frac{11}{4}M,\frac{13}{4}M\right]$ and equal to zero outside $\left(\frac{5}{2}M, \frac{7}{2}M\right)$. Clearly, to prove (\ref{Hardy}) it suffices to bound the expression
\begin{align} \label{Hardy2}
\int_0^{\tau} d\tilde{\tau} \int_{\Sigma_{\tilde{\tau}}} \chi^2 \left( (\partial_t \psi)^2 +  (\partial_r \psi)^2 + |\slashed{\nabla} \psi|^2\right) \left[\partial_r + \frac{2M+f(r)}{1-\frac{2M}{r}} \partial_t\right]\left(r-3M\right) 
\end{align}
by the right hand side of (\ref{Hardy}). This in turn follows after integrating the square bracket (which is proportional to $W$) by parts  and applying Cauchy's inequality with $\delta$ to the resulting spacetime terms.

\section{Final Comments} \label{sec:finalcomments}
We finally comment on the relation of (\ref{perturbed}) with the stability problem for black holes. We recall that for perturbations of the Kerr family the so-called Teukolsky equation \cite{teukolsky1973} governs the evolution of certain null curvature components in linearised theory. The Teukolsky equation can be viewed as a wave equation of the type (\ref{perturbed}) with \emph{large} first order terms preventing both the existence of a conserved energy and an ILED. Now, as crucially employed in the recent works  \cite{holzstabofschw, DHRteuk,Ma:2017yui2,PasqualottoMaxwell}, by suitable commutation(s), these first order terms can be entirely removed, the commuted quantities satisfying Regge-Wheeler-type equations for which a conserved energy and an ILED does exist.\footnote{The commutations are physical space manifestations of the classical Chandrasekhar transformations \cite{Chandrasekhar}.} To illustrate the above, while keeping the discussion in line with the elementary
case treated in this paper, we consider the Teukolsky equation for $s=1$, governing electromagnetic perturbations on Schwarzschild:
\begin{align} \label{Teukos1}
\left[TT - R^\star R^\star + \frac{1-\frac{2M}{r}}{r^2} (\Delta^{(1)}+1)\right] \alpha +\frac{2}{r} \left(1-\frac{3M}{r}\right) (T-R^\star) \alpha = 0 \, .
\end{align}
Here $\alpha$ is a spin-weighted function on $\mathcal{M}$ and $\Delta^{(1)}$ denotes the spin-$1$-weighted Laplacian, see \cite{DHRteuk}. The latter has eigenvalues $\ell(\ell+1)-1$ with $\ell \geq 1$, hence the square bracket should be viewed as the operator $(1-\frac{2M}{r})\mathcal{P}_s$ introduced in (\ref{psintro}), see footnote \ref{ftr}.
Commuting (\ref{Teukos1}) with the vectorfield $Z= \frac{r^2}{1-\frac{2M}{r}}(T-R^\star)$ results in the following Regge-Wheeler type equation for $Z\alpha$:
\begin{align}
\left[TT - R^\star R^\star + \frac{1-\frac{2M}{r}}{r^2} (\Delta^{(1)}+1)\right]Z\alpha = 0 \, .
\end{align}
Therefore, commuting with $Z$ has entirely removed the large \emph{degenerate} (at $r=3M$) first order term in (\ref{Teukos1}), and done so remarkably without producing any additional lower order derivatives of $\alpha$.

In contrast with this, the vectorfield $W=r\left(1-\frac{2M}{r}\right)^{-1/2}(R^\star -f(r)T)$ introduced in (\ref{Wdef}) generates upon commutation with the operator $\mathcal{P}_s$ a favourable\footnote{that is producing a spacetime term of the right sign when the $T$-energy estimate is applied.} \emph{non-degenerate} first order term on the right hand side (\ref{regcommuted}), however, at the cost of introducing first order error-terms in $\alpha$. 

We conclude that even if the vectorfield $W$ cannot replace the transformation theory described above, it does provide a more robust perspective on it. For instance, when generalising to Kerr, the analogue of $Z$ which ensures complete absence of first order terms in the transformed equation will (in Boyer-Lindquist coordinates) be of the form $\frac{(r^2+a^2)^2}{r^2-2Mr+a^2}(\partial_t- \partial_{r^\star} + \frac{a}{r^2+a^2}\partial_{\phi})$, see \cite{DHRteuk}. Now while not admitting the same amount of cancellations, using instead say, $\frac{r^4}{r^2-2Mr+a^2}(\partial_t- \partial_{r^\star} + \frac{a}{r^2+a^2}\partial_{\phi})$, will (at least in the case of small angular momentum) also result in a set of equations for which decay can be proven using the techniques introduced in this paper.

\section{Acknowledgements}
The ideas leading to this paper arose from discussions of the first author with Igor Rodnianski in 2017. 
The authors also thank Igor Rodnianski for comments on the manuscript. Support through ERC Consolidator Grant 772249 is gratefully acknowledged. 

\bibliographystyle{hacm}
\bibliography{wavefirstorder2}

\begin{thebibliography}{10}

\bibitem{AndBlue}
{\sc Andersson, L., and Blue, P.}
\newblock Hidden symmetries and decay for the wave equation on the {K}err
  spacetime.
\newblock {\em Ann. of Math. (2) 182}, 3 (2015), 787--853.

\bibitem{Angelopoulos:2016moe}
{\sc Angelopoulos, Y., Aretakis, S., and Gajic, D.}
\newblock {A vector field approach to almost-sharp decay for the wave equation
  on spherically symmetric, stationary spacetimes}.
\newblock {\em Ann.~PDE 4 (15)\/} (2018), arXiv:1612.01565.

\bibitem{Angelopoulos:2016wcv}
{\sc Angelopoulos, Y., Aretakis, S., and Gajic, D.}
\newblock {Late-time asymptotics for the wave equation on spherically
  symmetric, stationary spacetimes}.
\newblock {\em Adv. Math. 323\/} (2018), 529--621, arXiv:1612.01566.

\bibitem{Aretakis}
{\sc Aretakis, S.}
\newblock Decay of axisymmetric solutions of the wave equation on extreme
  {K}err backgrounds.
\newblock {\em J. Funct. Anal. 263}, 9 (2012), 2770--2831.

\bibitem{MR1972492}
{\sc Blue, P., and Soffer, A.}
\newblock Semilinear wave equations on the {S}chwarzschild manifold. {I}.
  {L}ocal decay estimates.
\newblock {\em Adv. Differential Equations 8}, 5 (2003), 595--614.
\newblock [Erratum: arXiv:gr-qc/0608073].

\bibitem{Sterbenz}
{\sc Blue, P., and Sterbenz, J.}
\newblock Uniform decay of local energy and the semi-linear wave equation on
  {S}chwarzschild space.
\newblock {\em Commun. Math. Phys. 268 (2)\/} (2006), 481--504.

\bibitem{Chandrasekhar}
{\sc Chandrasekhar, S.}
\newblock {\em The Mathematical Theory of Black Holes}, 3rd~ed.
\newblock Oxford University Press, Oxford, 1992.

\bibitem{DHRteuk}
{\sc Dafermos, M., Holzegel, G., and Rodnianski, I.}
\newblock Boundedness and decay for the {T}eukolsky equation on {K}err
  spacetimes {I}: the case $|a|\ll m$.
\newblock {\em Ann.~PDE 5 (2)\/} (2019), arXiv:1711.07944.

\bibitem{holzstabofschw}
{\sc Dafermos, M., Holzegel, G., and Rodnianski, I.}
\newblock {The linear stability of the Schwarzschild solution to gravitational
  perturbations}.
\newblock {\em Acta Math. 222\/} (2019), 1--214, arXiv:1601.06467.

\bibitem{DafRodnew}
{\sc Dafermos, M., and Rodnianski, I.}
\newblock A new physical-space approach to decay for the wave equation with
  applications to black hole spacetimes.
\newblock In {\em XVIth International Congress on Mathematical Physics},
  P.~Exner, Ed. World Scientific, London, 2009, pp.~421--433, arXiv:0910.4957.

\bibitem{DafRod2}
{\sc Dafermos, M., and Rodnianski, I.}
\newblock {The red-shift effect and radiation decay on black hole spacetimes}.
\newblock {\em Comm. Pure Appl. Math. 62\/} (2009), 859--919, gr-qc/0512119.

\bibitem{DafRodsmalla}
{\sc Dafermos, M., and Rodnianski, I.}
\newblock Decay for solutions of the wave equation on {K}err exterior
  spacetimes {I}-{II}: The cases $|a| \ll m$ or axisymmetry.
\newblock {\em arXiv:1010.5132, preprint\/} (2010).

\bibitem{Mihalisnotes}
{\sc Dafermos, M., and Rodnianski, I.}
\newblock {Lectures on black holes and linear waves}.
\newblock In {\em Evolution equations, Clay Mathematics Proceedings, Vol. 17}.
  Amer. Math. Soc., Providence, RI, 2013, pp.~97--205, arXiv:0811.0354.

\bibitem{partiii}
{\sc Dafermos, M., Rodnianski, I., and Shlapentokh-Rothman, Y.}
\newblock Decay for solutions of the wave equation on {K}err exterior
  spacetimes {III}: {T}he full subextremal case {$|a|<M$}.
\newblock {\em Ann. of Math. (2) 183}, 3 (2016), 787--913.

\bibitem{Dyatlovnh}
{\sc Dyatlov, S.}
\newblock {Spectral gaps for normally hyperbolic trapping}.
\newblock {\em Annales Inst. Fourier 66}, 1 (2016), 55--82, arXiv:1403.6401.

\bibitem{Hintznh}
{\sc Hintz, P.}
\newblock {Normally hyperbolic trapping on asymptotically stationary
  spacetimes}.
\newblock arXiv:1811.07843.

\bibitem{Hintz:2020roc}
{\sc Hintz, P.}
\newblock {A sharp version of Price's law for wave decay on asymptotically flat
  spacetimes}.
\newblock arXiv:2004.01664.

\bibitem{Lindbladglobal2016}
{\sc Lindblad, H., and Tohaneanu, M.}
\newblock {Global existence for quasilinear wave equations close to
  Schwarzschild}.
\newblock {\em Communications in Partial Differential Equations 43}, 6 (2018),
  893--944, arXiv:1610.00674.

\bibitem{Lindblad:2020bkn}
{\sc Lindblad, H., and Tohaneanu, M.}
\newblock {A local energy estimate for wave equations on metrics asymptotically
  close to Kerr}.
\newblock arXiv:2004.05664.

\bibitem{Ma:2017yui2}
{\sc Ma, S.}
\newblock {Uniform energy bound and Morawetz estimate for extreme components of
  spin fields in the exterior of a slowly rotating Kerr black hole \emph{II}:
  linearized gravity}.
\newblock {\em Comm.~Math.~Phys.\/} (2020), arXiv:1708.07385.

\bibitem{tohaneanu}
{\sc Metcalfe, J., Tataru, D., and Tohaneanu, M.}
\newblock Price's law on nonstationary space-times.
\newblock {\em Adv.~Math. 230}, 3 (2012), 995--1028.

\bibitem{PasqualottoMaxwell}
{\sc Pasqualotto, F.}
\newblock {The spin $\pm$1 Teukolsky equations and the Maxwell system on
  Schwarzschild}.
\newblock {\em Annales Henri Poincare 20}, 4 (2019), 1263--1323,
  arXiv:1612.07244.

\bibitem{sbierskithesis}
{\sc Sbierski, J.}
\newblock {\em On the initial value problem in general relativity and wave
  propagation in black-hole spacetimes}.
\newblock PhD thesis, University of Cambridge, Cambridge, 2014.

\bibitem{SbierskiGauss}
{\sc Sbierski, J.}
\newblock Characterisation of the energy of {G}aussian beams on {L}orentzian
  manifolds: with applications to black hole spacetimes.
\newblock {\em Anal. PDE 8}, 6 (2015), 1379--1420.

\bibitem{Toha2}
{\sc Tataru, D., and Tohaneanu, M.}
\newblock {Local energy estimate on Kerr black hole backgrounds}.
\newblock {\em Int. Math. Res. Not. 2011\/} (2011), 248--292, arXiv:0810.5766.

\bibitem{teukolsky1973}
{\sc Teukolsky, S.~A.}
\newblock Perturbations of a rotating black hole. {I}. {F}undamental equations
  for gravitational, electromagnetic, and neutrino-field perturbations.
\newblock {\em Astrophysical J. 185\/} (1973), 635--648.

\bibitem{Wunsch}
{\sc Wunsch, J., and Zworski, M.}
\newblock Resolvent estimates for normally hyperbolic trapped sets.
\newblock {\em Annales Henri Poincar{\'e} 12}, 7 (2011), 1349.

\end{thebibliography}

\end{document}